\documentclass[aps,pre,twocolumn,showpacs,preprintnumbers,amsmath,amssymb,floatfix]{revtex4}

\usepackage{graphicx}
\usepackage{dcolumn}
\usepackage{bm}
\begin{document}
\graphicspath{{figures/}{}}

\preprint{Phys.\ Rev.\ E 83, 047301 (2011) --- DOI : 10.1103/PhysRevE.83.047301 --- More information: http://fisica.unav.es/gmm/\ \ }
\title{1-D Cluster Array at the Three Phase Contact Line in Diluted Colloids Subjected to A.C.\ Electric Fields}

\author{M.\ Pichumani}
\author{M.\ Giuliani}
\altaffiliation{Present address: Dept.\ of Phys., University of Guelph, Canada}
\author{W.\ Gonz\'alez--Vi\~nas}%
\email{wens@fisica.unav.es}
\affiliation{Dept.\ of Phys.\ and Appl.\ Math., University of Navarra, Irunlarrea s/n, 31080 Pamplona, Spain}%

\date{March 18, 2011}
\begin{abstract}
Colloidal particles provide an efficient mean of building multiple scale structured materials from colloidal dispersions. In this Brief Report, we account for experimental evidence on the formation of a colloidal cluster array at a three-phase contact line. We study the influence of low frequency external alternating electric fields on a diluted colloidal dispersion opened to the air. We focus on the cluster formation and their evolution in the meniscus by measuring characteristic times and lengths. We observe that the clusters are separated by a well-defined length and that, in our experimental conditions, they survive between five a fifteen minutes. These new results could be of technological relevance in building tailored colloidal structures in non-patterned substrates.
\end{abstract}
\pacs{47.57.J-,81.16.Rf,47.20.-k,}
\maketitle


Non equilibrium transitions by external forces in colloidal systems gained attention among research groups \cite{trau1996,arcos2008,yethiraj2004,gong2003,tao2009,yethiraj2007,maxiepj2010} lately. Normally, these kind of transitions are more complex in nature because their interrelated parameters are affected by the external forces. These transitions occur when the state of the system becomes unstable as the control parameters are varied and usually lead to pattern formation \cite{cross1993,denis1994}. Flow dynamics plays an important role in the evolution of the system by transporting colloidal particles in the transitions \cite{giuliani2010}. The dynamic response of the colloidal dispersion to the applied external fields describes the basic functionality of the system and their inter-relative behavior (between clusters, and between clusters and substrates). Previous works \cite{zhang2009,ristenpart2008,ajay2005} study the influence of external fields on bulk colloidal dispersions. Here, we focus our attention in the meniscus of a colloid with a free surface, rather than in the bulk dispersion. Due to minimal evaporation in our setup, the contact line does not recede significantly during the measurement time. Consequently, the cluster formation is not related to a fingering instability in the contact line \cite{debruyn1992,brenner1993,diez2001}. Also, we explore the effect of weak external  ac electric fields in this region. On comparing with the surface and other kind of instabilities \cite{yue1994,schaffer2001}, the one which we obtain is mainly comprised of two phenomena: (a) Electrokinetic (b) Electrowetting \cite{kwan2002,mugele2005}. The possible effect from electrokinetic phenomena (mainly electrophoresis) on the dispersion is the formation of a concentration gradient (maintaining  high concentration of particles near the contact line which is greater than the initial concentration). On the other hand, AC electrowetting induces local flows in the meniscus which may form clusters. Both contributions set a threshold for the forcing in order to obtain clusters. In this work, we investigate the effect of external weak alternating electric fields on a diluted colloidal dispersion. In these conditions, the interparticle interactions may play a critical role on transport properties \cite{yamanaka1990,yamanaka1991,ashok2009}. We focus in a region of the parameters space where clusters form. We study three relevant aspects: (i) The response of the clusters to the applied field. (ii) The evolution of the system by measuring the time required for the cluster formation. (iii)~The 1-D spatial distribution of the clusters and their evolution in time. The aim of this work is to get insights on the mechanisms involved in colloidal cluster formation and their evolution.

\begin{figure}[!ht]
\includegraphics[width=8.5cm]{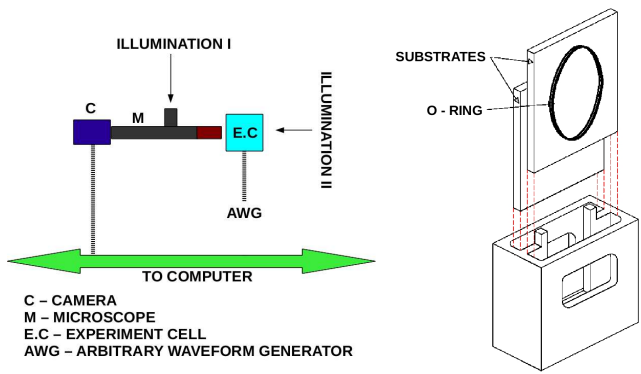}
\caption{\label{exptsetup}(Color online) Sketch of the experimental setup (left) and of the cell (right).}
\end{figure}

The experimental setup and the cell is shown in figure \ref{exptsetup}. The substrates used for all the experiments are of standard glass (of size 1.1x17x18 mm$^3$) in which one of the sides (17x18 mm$^2$) is coated with a thin conductive layer of Indium Tin Oxide (ITO). They are placed vertically in a Teflon cell (Fig.~\ref{exptsetup} - Right). The substrates are placed in such a way that, the conductive side is in contact with the suspension. The thickness of the colloidal suspension in between the two substrates is of 1~mm. To apply electric fields, we place two ceramic spacers (one side of each ceramic spacer is made conductive) between the substrates. Rubber O-rings are used to ensure that the cell along with the substrates and spacers are held tightly. All the experiments are performed at room temperature. We use a colloidal dispersion of spherical polystyrene particles of diameter 1.3 $\mu$m (polydispersity is 0.039). These particles are suspended in water and their surface charge is -7~$\mu$C/cm$^2$. The dispersion is acquired from Dr.\ Paulke at Fraunhofer-IAP, Germany. We dilute this stock dispersion to 0.5\% (w/w) concentration using ultra pure water.

\begin{figure}[!ht]
\includegraphics[width=8.5cm]{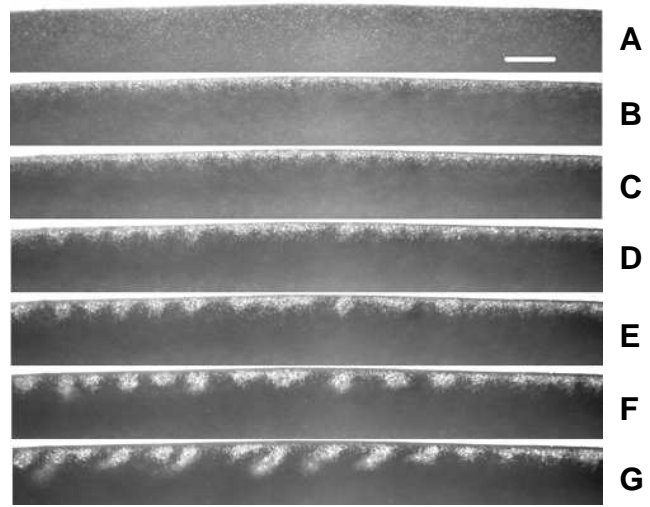}
\caption{\label{snap}Snapshots of the contact line region (front view) at 1.8~Vpp and 0.8~Hz. The bright regions are particles or clusters. The frames are captured at intervals of 1 minute and the scale bar is 50 $\mu$m. The stages (A-G) are explained in the text.}
\end{figure}

The substrates are cleaned with nitrogen blow and they are mounted in the cell. A subset of the substrates were prepared with more hydrophilic character in order to evaluate the influence of the equilibrium contact angle on the cluster formation. Electric fields (AC) of square wave form are applied perpendicularly to the conducting substrates. We apply low voltage (of the order of 1~Vpp/mm) and low frequency (of the order of 1~Hz). We observe the region near the contact line using a microscope with an objective of magnification 10X. This microscope is attached to a CMOS camera for capturing sequential frames (Fig.~\ref{snap}). The location of the contact line corresponds to the transition between the dark region (colloidal suspension) at the bottom and the white region (bare substrate) at the top. The area of observation corresponds to the central part of the contact line region with a field of view of 0.6 mm. However the observed phenomenology is common to the whole contact line (far from boundaries).


After the dispersion is loaded into the cell (Fig.~\ref{snap} A), we apply the electric field and images of the system are saved sequentially. Due to the electrophoretic effect the particles rhythmically get closer to the contact line. They are trapped in the meniscus and start to accumulate (Fig.~\ref{snap}, B and C). After some time, the clusters start to initiate (Fig.~\ref{snap} D) and they continue to grow (Fig.~\ref{snap}, E and F), until they stabilize their size. During the growing stage, clusters change their shape at the same frequency of the applied field. Finally, the clusters will age/dissociate (Fig.~\ref{snap} G). See movie in \cite{moviea}.

In order to obtain an initial picture about the system, we observe the response of particles to the applied field. From the stored frames, we construct a space time diagram, from which their frequency is obtained. We measured the response of the particles to the external forcing in two cases: (i) By keeping the frequency constant (1~Hz) and varying the amplitude. (ii) By keeping the amplitude constant (1.8~Vpp) and varying the frequency. We checked that for the range of applied amplitudes (from 1.8 to 2.4~Vpp) at a constant frequency of 1 Hz, the measured frequencies of the particles coincide with the applied one (1.00$\pm$0.06~Hz). In both cases (i and ii), we confirm that in our experimental conditions, the particles follow adiabatically the applied field.

\begin{figure}[!ht]
\includegraphics[width=8cm]{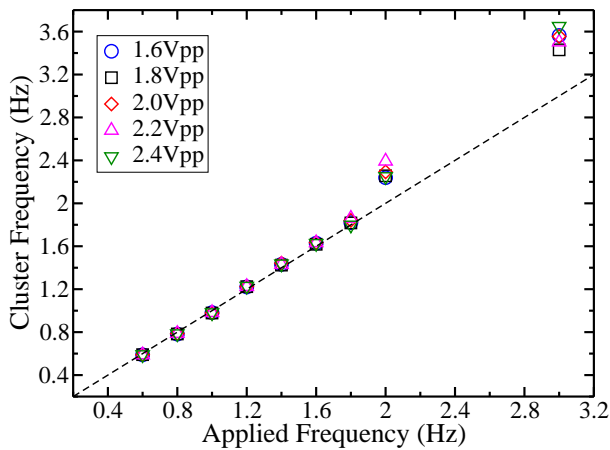}
\caption{\label{clusterfreq}(Color online) Frequency response of the clusters to the applied field. The dashed line is an indicative to the applied frequencies.}
\end{figure}

\begin{figure}[!ht]
\includegraphics[width=8cm]{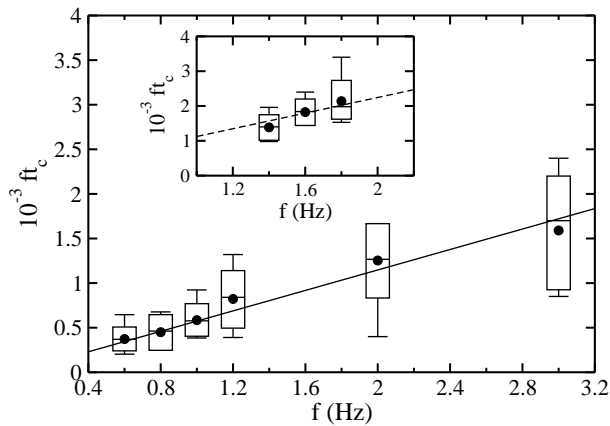}
\caption{\label{tfvsf}Non dimensional threshold time (ft$_c$) as a function of applied frequencies. Boxplots merge data for different strengths of the applied field. For each frequency, the mean non dimensional threshold time is represented by a filled circle. The lines are fits to a proportional law (solid corresponds to t$_{c}$ = 574~s and dashed to t$_{c}$ = 1122~s). Inset corresponds to experiments performed with more hydrophilic substrates.}
\end{figure}

Concerning the response of clusters to the applied field, we define a horizontal line in the cluster region for all the experimental frames. Along this line the brightness is captured, in such a way that the intensity varies when the clusters move and/or they change their intrinsic brightness. We construct the corresponding space-time diagram followed by a 2-D Fast Fourier Transformation (FFT) which gives a frequency related to the clusters (Fig.~\ref{clusterfreq}). In the case of applied low frequencies (0.6~Hz to 1.8~Hz), the clusters follow the forcing condition whereas the response is of higher values at applied high frequencies (2.0~Hz and 3.0~Hz). To study further the behavior of clusters and its relevant parameters (in correlating their formation with the applied field), we measured the temporal scale at which they form/evolve. To estimate this, we measured the threshold time (t$_c$) which we define as the time beyond which the clusters start to age/dissociate. The non dimensional threshold time (ft$_c$) as a function of applied frequencies (f) is shown in figure \ref{tfvsf} and it increases proportionally to the increasing applied frequencies. In a previous learning section, it was observed that this behavior does not depend, within the experimental error and in our experimental conditions, on the strength of the applied field.

\begin{figure}[!ht]
\includegraphics[width=8cm]{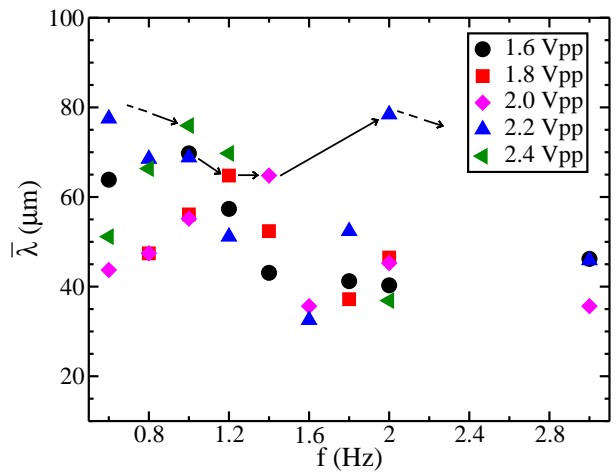}
\caption{\label{lamvsf}(Color online) Characteristic length ($\bar{\lambda}$) at the threshold time (t$_c$) as a function of applied frequencies. The arrows outline the sequence of the maxima occuring while increasing the strength of the field. The dashed line arrows correspond to the 2.2 Vpp to 2.4 Vpp transition (see text).}
\end{figure}

The clusters are spatially distributed and they are separated by a length $\lambda$. We measured the characteristic length at the threshold time t$_c$ (Fig.~\ref{lamvsf}). The characteristic length $\bar{\lambda}$ defines the mean spatial distance between the clusters calculated through the spatial auto-correlation function. If we consider for one applied amplitude (e.g.\ 2.4~Vpp in figure \ref{lamvsf}), the characteristic length evolves as we increase the frequency. For one particular frequency (e.g.\ at 1.0~Hz), a relative maximum in characteristic length is observed.

\begin{figure}[!ht]
\includegraphics[width=8.5cm]{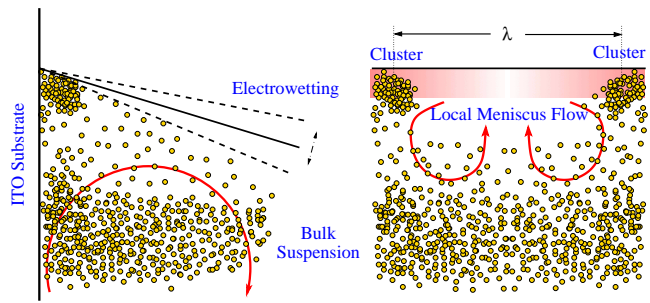}
\caption{\label{sketch}(Color online) Microscopic sketch of the region near the contact line. Left: Lateral view. Effect due to electrowetting has been enhanced for better view. Right: Front view. Local flows are indicated by red (curved) arrows and the corresponding transverse modulation in the contact angle is shown by the color (shade) gradient near the contact line.}
\end{figure}

On applying the electric field, the electrophoretic effect supplies particles to the meniscus much more efficiently than the Brownian motion does. As it is stated above, surface forces trap particles in the meniscus. Secondly, AC electrowetting \cite{kwan2002} generates a flow \cite{ko2008,lee2009} near the contact line (due to the increase and decrease in the contact angle) that may become unstable leading to a transverse modulation which would be responsible for the cluster formation. These effects are sketched in figure \ref{sketch}.

Initially, when the electric field is turned on, the particles get accumulated like a band near the contact line (Fig.~\ref{snap}, B and C). At later stages, these particles initiate the formation of clusters and they evolve. Similarly as in the case of the response of particles, the clusters follow the force at lower order of applied frequencies (Fig.~\ref{clusterfreq} from 0.6~Hz to 1.8~Hz) by moving as a whole up and down (perpendicularly to the applied field). Thus, the clusters can follow the variations in the contact angle, as it is observed in the experiments. But contrarily, the measured frequencies of the clusters is of higher values for applied high frequencies. For these frequencies, the rapid variations in the contact angle make these clusters unable to follow and they do not move, due to their inertia. However, the particles rearrange inside the clusters. This could lead to high frequency variations in the measured light intensity without global movement of the clusters (see movie in \cite{movieb}). 

The long temporal scale behavior of the system can be characterized by measuring the time at which the clusters evolve significantly. The proportional increase in the non dimensional time (Fig.~\ref{tfvsf}) indicates that this global evolution of the system (time required for the clusters to age/dissociate) is weakly dependent on the forcing condition. The measured higher non dimensional time in figure \ref{tfvsf} (inset) is due to the character of the substrate (less hydrophobic) and the aging is due to a global scale flow which disassembles the clusters (see movie in \cite{moviea}).

\begin{figure}[!ht]
\includegraphics[width=8cm]{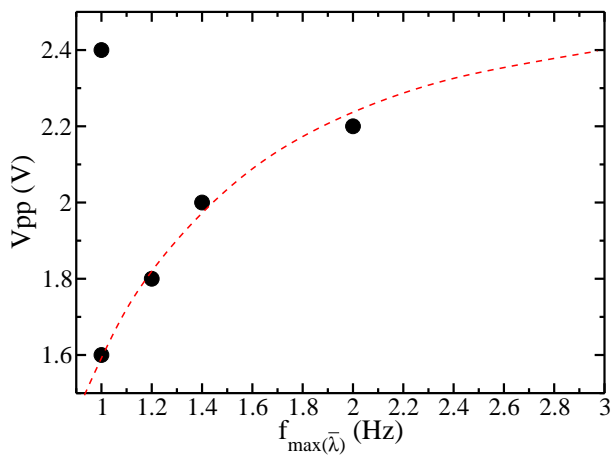}
\caption{\label{vppvsfmax}(Color online) Correlation between applied amplitude and formation of clusters for relative maximum in characteristic length $\bar{\lambda}$. The dashed line is a guide to the eye.}
\end{figure}

An instability on the flows (which are induced by the AC electrowetting) gives rise to transverse local flows and to a modulation of the varying contact angle (Fig.~\ref{sketch} - right). This local flows are advection-like closed flows. We call the recurrence time, the one taken by a volume element to complete a cycle. The characteristic length of this closed flows corresponds to the mean distance between clusters ($\bar{\lambda}$). The characteristic length changes as we increase the applied frequency (see arrows in figure~\ref{lamvsf}). For a certain frequency, a relative maximum in the characteristic length is observed. This frequency depends on the recurrence time, and thus on the strength of the local flows. To have this condition (relative maximum in characteristic length) for increasing applied frequencies, the amplitude has to be increased (Fig.~\ref{vppvsfmax}). The recurrence time is limited in order to have a regular flow. Consequently, for a high applied amplitude (e.g.\ 2.4~Vpp) the relative maximum in characteristic length is obtained at a lower applied frequency, that corresponds to a multiple of the expected recurrence time (see dashed line arrows in figure~\ref{lamvsf}, and figure~\ref{vppvsfmax}). This may indicate that the instability is parametric \cite{poulin2003}.

In conclusion, the applied alternating field generates flows near the contact line and clusters are formed. The clusters are separated by a well-defined characteristic length ($\bar{\lambda}$). The clusters are located in the meniscus and they survive in our experimental conditions between five and fifteen minutes. The response of the contact angle depends on the applied frequencies and it is measured through the frequency response of the clusters. Also, the applied field controls the flows in the meniscus (through the variation of contact angle). These results will be complemented in further work with the characterization of the instability (thresholds and type of bifurcation) and of the phase space. It is crucial to understand the underlying fluid dynamics in order to control the periodicity of the cluster array, which is very relevant in applications where colloidal multi-scale structures are needed. For this, both new experiments and models are required. Our experimental set-up and conditions allow to build this kind of colloidal deposits at low cost and in a straightforward manner. Although we have checked that electroosmosis \cite{trau1997,solomentsev1997} does not play an important role in this experiment, more research has to be performed to determine quantitatively this issue.

\begin{acknowledgments}
This work is partly supported by the Spanish Government Contract  No.\ FIS2008-01126 and by the Gobierno de Navarra (Departamento de Educaci\'on). MP and MG acknowledge financial support from the ``Asociaci\'on de Amigos de la Universidad de Navarra''
\end{acknowledgments}

\bibliography{ac_arxiv}

\begin{thebibliography}{30}
\expandafter\ifx\csname natexlab\endcsname\relax\def\natexlab#1{#1}\fi
\expandafter\ifx\csname bibnamefont\endcsname\relax
  \def\bibnamefont#1{#1}\fi
\expandafter\ifx\csname bibfnamefont\endcsname\relax
  \def\bibfnamefont#1{#1}\fi
\expandafter\ifx\csname citenamefont\endcsname\relax
  \def\citenamefont#1{#1}\fi
\expandafter\ifx\csname url\endcsname\relax
  \def\url#1{\texttt{#1}}\fi
\expandafter\ifx\csname urlprefix\endcsname\relax\def\urlprefix{URL }\fi
\providecommand{\bibinfo}[2]{#2}
\providecommand{\eprint}[2][]{\url{#2}}

\bibitem[{\citenamefont{Trau et~al.}(1996)\citenamefont{Trau, Saville, and
  Aksay}}]{trau1996}
\bibinfo{author}{\bibfnamefont{M.}~\bibnamefont{Trau}},
  \bibinfo{author}{\bibfnamefont{D.~A.} \bibnamefont{Saville}},
  \bibnamefont{and} \bibinfo{author}{\bibfnamefont{I.~A.} \bibnamefont{Aksay}},
  \bibinfo{journal}{Science} \textbf{\bibinfo{volume}{272}},
  \bibinfo{pages}{706} (\bibinfo{year}{1996}).

\bibitem[{\citenamefont{Arcos et~al.}(2008)\citenamefont{Arcos, Kumar,
  Gonz{\'a}lez-Vi{\~n}as, Sirera, Poduska, and Yethiraj}}]{arcos2008}
\bibinfo{author}{\bibfnamefont{C.}~\bibnamefont{Arcos}},
  \bibinfo{author}{\bibfnamefont{K.}~\bibnamefont{Kumar}},
  \bibinfo{author}{\bibfnamefont{W.}~\bibnamefont{Gonz{\'a}lez-Vi{\~n}as}},
  \bibinfo{author}{\bibfnamefont{R.}~\bibnamefont{Sirera}},
  \bibinfo{author}{\bibfnamefont{K.~M.} \bibnamefont{Poduska}},
  \bibnamefont{and} \bibinfo{author}{\bibfnamefont{A.}~\bibnamefont{Yethiraj}},
  \bibinfo{journal}{Phys. Rev. E} \textbf{\bibinfo{volume}{77}},
  \bibinfo{pages}{050402(R)} (\bibinfo{year}{2008}).

\bibitem[{\citenamefont{Yethiraj et~al.}(2004)\citenamefont{Yethiraj, Wouterse,
  Groh, and van Blaaderen}}]{yethiraj2004}
\bibinfo{author}{\bibfnamefont{A.}~\bibnamefont{Yethiraj}},
  \bibinfo{author}{\bibfnamefont{A.}~\bibnamefont{Wouterse}},
  \bibinfo{author}{\bibfnamefont{B.}~\bibnamefont{Groh}}, \bibnamefont{and}
  \bibinfo{author}{\bibfnamefont{A.}~\bibnamefont{van Blaaderen}},
  \bibinfo{journal}{Phys. Rev. Lett.} \textbf{\bibinfo{volume}{92}},
  \bibinfo{pages}{058301} (\bibinfo{year}{2004}).

\bibitem[{\citenamefont{Gong et~al.}(2003)\citenamefont{Gong, Wu, and
  Marr}}]{gong2003}
\bibinfo{author}{\bibfnamefont{T.}~\bibnamefont{Gong}},
  \bibinfo{author}{\bibfnamefont{D.~T.} \bibnamefont{Wu}}, \bibnamefont{and}
  \bibinfo{author}{\bibfnamefont{D.~W.~M.} \bibnamefont{Marr}},
  \bibinfo{journal}{Langmuir} \textbf{\bibinfo{volume}{19}},
  \bibinfo{pages}{5967} (\bibinfo{year}{2003}).

\bibitem[{\citenamefont{H-Tao et~al.}(2009)\citenamefont{H-Tao, Ming, Y-Xian,
  and Ping}}]{tao2009}
\bibinfo{author}{\bibfnamefont{Y.}~\bibnamefont{H-Tao}},
  \bibinfo{author}{\bibfnamefont{W.}~\bibnamefont{Ming}},
  \bibinfo{author}{\bibfnamefont{G.}~\bibnamefont{Y-Xian}}, \bibnamefont{and}
  \bibinfo{author}{\bibfnamefont{Y.}~\bibnamefont{Ping}},
  \bibinfo{journal}{Chinese Phys. B} \textbf{\bibinfo{volume}{18}},
  \bibinfo{pages}{2389} (\bibinfo{year}{2009}).

\bibitem[{\citenamefont{Yethiraj}(2007)}]{yethiraj2007}
\bibinfo{author}{\bibfnamefont{A.}~\bibnamefont{Yethiraj}},
  \bibinfo{journal}{Soft Matter} \textbf{\bibinfo{volume}{3}},
  \bibinfo{pages}{1099} (\bibinfo{year}{2007}).

\bibitem[{\citenamefont{Giuliani et~al.}(2011)\citenamefont{Giuliani,
  Pichumani, and Gonz{\'a}lez-Vi{\~n}as}}]{maxiepj2010}
\bibinfo{author}{\bibfnamefont{M.}~\bibnamefont{Giuliani}},
  \bibinfo{author}{\bibfnamefont{M.}~\bibnamefont{Pichumani}},
  \bibnamefont{and}
  \bibinfo{author}{\bibfnamefont{W.}~\bibnamefont{Gonz{\'a}lez-Vi{\~n}as}},
  \bibinfo{journal}{Eur. Phys. J. Special Topics}
  \textbf{\bibinfo{volume}{192}}, \bibinfo{pages}{121} (\bibinfo{year}{2011}).

\bibitem[{\citenamefont{Cross and Hohenberg}(1993)}]{cross1993}
\bibinfo{author}{\bibfnamefont{M.}~\bibnamefont{Cross}} \bibnamefont{and}
  \bibinfo{author}{\bibfnamefont{P.~C.} \bibnamefont{Hohenberg}},
  \bibinfo{journal}{Rev. Mod. Phys} \textbf{\bibinfo{volume}{65}}
  (\bibinfo{year}{1993}).

\bibitem[{\citenamefont{Wirtz and Fermingier}(1994)}]{denis1994}
\bibinfo{author}{\bibfnamefont{D.}~\bibnamefont{Wirtz}} \bibnamefont{and}
  \bibinfo{author}{\bibfnamefont{M.}~\bibnamefont{Fermingier}},
  \bibinfo{journal}{Phys. Rev. Lett.} \textbf{\bibinfo{volume}{72}},
  \bibinfo{pages}{2294} (\bibinfo{year}{1994}).

\bibitem[{\citenamefont{Giuliani et~al.}(2010)\citenamefont{Giuliani,
  Gonz{\'a}lez-Vi{\~n}as, Poduska, and Yethiraj}}]{giuliani2010}
\bibinfo{author}{\bibfnamefont{M.}~\bibnamefont{Giuliani}},
  \bibinfo{author}{\bibfnamefont{W.}~\bibnamefont{Gonz{\'a}lez-Vi{\~n}as}},
  \bibinfo{author}{\bibfnamefont{K.}~\bibnamefont{Poduska}}, \bibnamefont{and}
  \bibinfo{author}{\bibfnamefont{A.}~\bibnamefont{Yethiraj}},
  \bibinfo{journal}{J. Phys. Chem. Lett.} \textbf{\bibinfo{volume}{1}},
  \bibinfo{pages}{1481} (\bibinfo{year}{2010}).

\bibitem[{\citenamefont{Zhang and Liu}(2009)}]{zhang2009}
\bibinfo{author}{\bibfnamefont{K.~Q.} \bibnamefont{Zhang}} \bibnamefont{and}
  \bibinfo{author}{\bibfnamefont{X.~Y.} \bibnamefont{Liu}},
  \bibinfo{journal}{J. Chem. Phys.} \textbf{\bibinfo{volume}{130}},
  \bibinfo{pages}{184901} (\bibinfo{year}{2009}).

\bibitem[{\citenamefont{Ristenpart et~al.}(2008)\citenamefont{Ristenpart,
  Jiang, Slowik, Punckt, Saville, and Aksay}}]{ristenpart2008}
\bibinfo{author}{\bibfnamefont{W.~D.} \bibnamefont{Ristenpart}},
  \bibinfo{author}{\bibfnamefont{P.}~\bibnamefont{Jiang}},
  \bibinfo{author}{\bibfnamefont{M.~A.} \bibnamefont{Slowik}},
  \bibinfo{author}{\bibfnamefont{C.}~\bibnamefont{Punckt}},
  \bibinfo{author}{\bibfnamefont{D.~A.} \bibnamefont{Saville}},
  \bibnamefont{and} \bibinfo{author}{\bibfnamefont{I.~A.} \bibnamefont{Aksay}},
  \bibinfo{journal}{Langmuir} \textbf{\bibinfo{volume}{24}},
  \bibinfo{pages}{12172} (\bibinfo{year}{2008}).

\bibitem[{\citenamefont{Negi et~al.}(2005)\citenamefont{Negi, Sengupta, and
  Sood}}]{ajay2005}
\bibinfo{author}{\bibfnamefont{A.~S.} \bibnamefont{Negi}},
  \bibinfo{author}{\bibfnamefont{K.}~\bibnamefont{Sengupta}}, \bibnamefont{and}
  \bibinfo{author}{\bibfnamefont{A.~K.} \bibnamefont{Sood}},
  \bibinfo{journal}{Langmuir} \textbf{\bibinfo{volume}{21}},
  \bibinfo{pages}{11623} (\bibinfo{year}{2005}).

\bibitem[{\citenamefont{de~Bruyn}(1992)}]{debruyn1992}
\bibinfo{author}{\bibfnamefont{J.}~\bibnamefont{de~Bruyn}},
  \bibinfo{journal}{Phys. Rev. A} \textbf{\bibinfo{volume}{46}},
  \bibinfo{pages}{R4500} (\bibinfo{year}{1992}).

\bibitem[{\citenamefont{Brenner}(1993)}]{brenner1993}
\bibinfo{author}{\bibfnamefont{M.}~\bibnamefont{Brenner}},
  \bibinfo{journal}{Phys. Rev. E} \textbf{\bibinfo{volume}{47}},
  \bibinfo{pages}{4597} (\bibinfo{year}{1993}).

\bibitem[{\citenamefont{Diez and Kondic}(2001)}]{diez2001}
\bibinfo{author}{\bibfnamefont{J.}~\bibnamefont{Diez}} \bibnamefont{and}
  \bibinfo{author}{\bibfnamefont{L.}~\bibnamefont{Kondic}},
  \bibinfo{journal}{Phys. Rev. Lett.} \textbf{\bibinfo{volume}{86}},
  \bibinfo{pages}{632} (\bibinfo{year}{2001}).

\bibitem[{\citenamefont{Hu et~al.}(1994)\citenamefont{Hu, Glass, Griffith, and
  Fraden}}]{yue1994}
\bibinfo{author}{\bibfnamefont{Y.}~\bibnamefont{Hu}},
  \bibinfo{author}{\bibfnamefont{J.~L.} \bibnamefont{Glass}},
  \bibinfo{author}{\bibfnamefont{A.~E.} \bibnamefont{Griffith}},
  \bibnamefont{and} \bibinfo{author}{\bibfnamefont{S.}~\bibnamefont{Fraden}},
  \bibinfo{journal}{J. Chem. Phys.} \textbf{\bibinfo{volume}{100}},
  \bibinfo{pages}{4674} (\bibinfo{year}{1994}).

\bibitem[{\citenamefont{Sch{\"a}ffer et~al.}(2001)\citenamefont{Sch{\"a}ffer,
  Thurn-Albrecht, Russell, and Steiner}}]{schaffer2001}
\bibinfo{author}{\bibfnamefont{E.}~\bibnamefont{Sch{\"a}ffer}},
  \bibinfo{author}{\bibfnamefont{T.}~\bibnamefont{Thurn-Albrecht}},
  \bibinfo{author}{\bibfnamefont{T.~P.} \bibnamefont{Russell}},
  \bibnamefont{and} \bibinfo{author}{\bibfnamefont{U.}~\bibnamefont{Steiner}},
  \bibinfo{journal}{Europhys. Lett} \textbf{\bibinfo{volume}{53}},
  \bibinfo{pages}{518} (\bibinfo{year}{2001}).

\bibitem[{\citenamefont{Kang}(2002)}]{kwan2002}
\bibinfo{author}{\bibfnamefont{K.~H.} \bibnamefont{Kang}},
  \bibinfo{journal}{Langmuir} \textbf{\bibinfo{volume}{18}},
  \bibinfo{pages}{10318} (\bibinfo{year}{2002}).

\bibitem[{\citenamefont{Mugele et~al.}(2005)\citenamefont{Mugele, Klingner,
  Buehrle, Steinhauser, and Herminghaus}}]{mugele2005}
\bibinfo{author}{\bibfnamefont{F.}~\bibnamefont{Mugele}},
  \bibinfo{author}{\bibfnamefont{A.}~\bibnamefont{Klingner}},
  \bibinfo{author}{\bibfnamefont{J.}~\bibnamefont{Buehrle}},
  \bibinfo{author}{\bibfnamefont{D.}~\bibnamefont{Steinhauser}},
  \bibnamefont{and}
  \bibinfo{author}{\bibfnamefont{S.}~\bibnamefont{Herminghaus}},
  \bibinfo{journal}{J. Phys.: Condens. Matter} \textbf{\bibinfo{volume}{17}},
  \bibinfo{pages}{S559} (\bibinfo{year}{2005}).

\bibitem[{\citenamefont{Yamanaka et~al.}(1990)\citenamefont{Yamanaka, Matsouka,
  Kitano, and Ise}}]{yamanaka1990}
\bibinfo{author}{\bibfnamefont{J.}~\bibnamefont{Yamanaka}},
  \bibinfo{author}{\bibfnamefont{H.}~\bibnamefont{Matsouka}},
  \bibinfo{author}{\bibfnamefont{H.}~\bibnamefont{Kitano}}, \bibnamefont{and}
  \bibinfo{author}{\bibfnamefont{N.}~\bibnamefont{Ise}}, \bibinfo{journal}{J.
  Colloid Interface Sci.} \textbf{\bibinfo{volume}{134}}, \bibinfo{pages}{92}
  (\bibinfo{year}{1990}).

\bibitem[{\citenamefont{Yamanaka et~al.}(1991)\citenamefont{Yamanaka, Matsouka,
  Kitano, Ise, Yamaguchi, Saeki, and Tsubokawa}}]{yamanaka1991}
\bibinfo{author}{\bibfnamefont{J.}~\bibnamefont{Yamanaka}},
  \bibinfo{author}{\bibfnamefont{H.}~\bibnamefont{Matsouka}},
  \bibinfo{author}{\bibfnamefont{H.}~\bibnamefont{Kitano}},
  \bibinfo{author}{\bibfnamefont{N.}~\bibnamefont{Ise}},
  \bibinfo{author}{\bibfnamefont{T.}~\bibnamefont{Yamaguchi}},
  \bibinfo{author}{\bibfnamefont{S.}~\bibnamefont{Saeki}}, \bibnamefont{and}
  \bibinfo{author}{\bibfnamefont{M.}~\bibnamefont{Tsubokawa}},
  \bibinfo{journal}{Langmuir} \textbf{\bibinfo{volume}{7}},
  \bibinfo{pages}{1928} (\bibinfo{year}{1991}).

\bibitem[{\citenamefont{Ashok and Muthukumar}(2009)}]{ashok2009}
\bibinfo{author}{\bibfnamefont{B.}~\bibnamefont{Ashok}} \bibnamefont{and}
  \bibinfo{author}{\bibfnamefont{M.}~\bibnamefont{Muthukumar}},
  \bibinfo{journal}{J. Phys. Chem. B} \textbf{\bibinfo{volume}{113}},
  \bibinfo{pages}{5736} (\bibinfo{year}{2009}).

\bibitem[{mov({\natexlab{a}})}]{moviea}
\bibinfo{note}{See EPAPS Document No. (to be inserted by the editor) for a
  movie which corresponds to figure 2 of this Letter.}

\bibitem[{\citenamefont{Ko et~al.}(2008)\citenamefont{Ko, Lee, and
  Kang}}]{ko2008}
\bibinfo{author}{\bibfnamefont{S.~H.} \bibnamefont{Ko}},
  \bibinfo{author}{\bibfnamefont{H.}~\bibnamefont{Lee}}, \bibnamefont{and}
  \bibinfo{author}{\bibfnamefont{K.~H.} \bibnamefont{Kang}},
  \bibinfo{journal}{Langmuir} \textbf{\bibinfo{volume}{24}},
  \bibinfo{pages}{1094} (\bibinfo{year}{2008}).

\bibitem[{\citenamefont{Lee et~al.}(2009)\citenamefont{Lee, Yun, Ko, and
  Kang}}]{lee2009}
\bibinfo{author}{\bibfnamefont{H.}~\bibnamefont{Lee}},
  \bibinfo{author}{\bibfnamefont{S.}~\bibnamefont{Yun}},
  \bibinfo{author}{\bibfnamefont{S.~H.} \bibnamefont{Ko}}, \bibnamefont{and}
  \bibinfo{author}{\bibfnamefont{K.~H.} \bibnamefont{Kang}},
  \bibinfo{journal}{Biomicrofluidics} \textbf{\bibinfo{volume}{3}},
  \bibinfo{pages}{044113} (\bibinfo{year}{2009}).

\bibitem[{mov({\natexlab{b}})}]{movieb}
\bibinfo{note}{See EPAPS Document No. (to be inserted by the editor) for a
  movie where the behavior of clusters at high frequency (3 Hz) can be seen.}

\bibitem[{\citenamefont{Poulin et~al.}(2003)\citenamefont{Poulin, Flierl, and
  Pedlosky}}]{poulin2003}
\bibinfo{author}{\bibfnamefont{F.}~\bibnamefont{Poulin}},
  \bibinfo{author}{\bibfnamefont{G.}~\bibnamefont{Flierl}}, \bibnamefont{and}
  \bibinfo{author}{\bibfnamefont{J.}~\bibnamefont{Pedlosky}},
  \bibinfo{journal}{J. Fluid Mech.} \textbf{\bibinfo{volume}{481}},
  \bibinfo{pages}{329} (\bibinfo{year}{2003}).

\bibitem[{\citenamefont{Trau et~al.}(1997)\citenamefont{Trau, Saville, and
  Aksay}}]{trau1997}
\bibinfo{author}{\bibfnamefont{M.}~\bibnamefont{Trau}},
  \bibinfo{author}{\bibfnamefont{D.}~\bibnamefont{Saville}}, \bibnamefont{and}
  \bibinfo{author}{\bibfnamefont{I.}~\bibnamefont{Aksay}},
  \bibinfo{journal}{Langmuir} \textbf{\bibinfo{volume}{13}},
  \bibinfo{pages}{6375} (\bibinfo{year}{1997}).

\bibitem[{\citenamefont{Solomentsev et~al.}(13)\citenamefont{Solomentsev,
  B{\"o}hmer, and Anderson}}]{solomentsev1997}
\bibinfo{author}{\bibfnamefont{Y.}~\bibnamefont{Solomentsev}},
  \bibinfo{author}{\bibfnamefont{M.}~\bibnamefont{B{\"o}hmer}},
  \bibnamefont{and} \bibinfo{author}{\bibfnamefont{J.}~\bibnamefont{Anderson}},
  \bibinfo{journal}{Langmuir} \textbf{\bibinfo{volume}{13}},
  \bibinfo{pages}{6058} (\bibinfo{year}{13}).

\end{thebibliography}

\end{document}